# Multi-Platform Generative Development of Component & Connector Systems using Model and Code Libraries

Jan Oliver Ringert[1,2][*], Bernhard Rumpe[1], and Andreas Wortmann[1]

[1] Software Engineering
RWTH Aachen University
http://www.se-rwth.de/
[2] School of Computer Science
Tel Aviv University
http://www.cs.tau.ac.il/

**Abstract.** Component-based software engineering aims to reduce software development effort by reusing established components as building blocks of complex systems. Defining components in general-purpose programming languages restricts their reuse to platforms supporting these languages and complicates component composition with implementation details. The vision of model-driven engineering is to reduce the gap between developer intention and implementation details by lifting abstract models to primary development artifacts and systematically transforming these into executable systems. For sufficiently complex systems the transformation from abstract models to platform-specific implementations requires augmentation with platform-specific components. We propose a model-driven mechanism to transform platform-independent logical component & connector architectures into platform-specific implementations combining model and code libraries. This mechanism allows to postpone commitment to a specific platform and thus increases reuse of software architectures and components.

## 1 Introduction

Component-based software engineering (CBSE) [16] ultimately aims to compose complex systems from off-the-shelf components. Usually, components are provided as general-purpose programming language (GPL) source code. This restricts reuse to certain platforms and requires domain experts to become programming experts. Model-driven engineering (MDE) pursues to reduce the conceptual gap [7] between domain and implementation concepts by describing software systems as abstract models. These models can be systematically transformed into implementations for potentially multiple target platforms. Component & connector (C&C) architecture description languages (ADLs) [15] are modeling languages with high potential to combine the benefits of MDE and CBSE. Software architectures can be modeled platform-independently, enriched with platform-specific information, and transformed into an implementation.

---

[*] J. O. Ringert acknowledges support from a postdoctoral Minerva Fellowship, funded by the German Federal Ministry for Education and Research.

We have developed the C&C ADL and framework MontiArcAutomaton [19,20] to facilitate MDE in robotics. MontiArcAutomaton supports the integration of the most suitable modeling languages and the composition and orchestration of independently developed code generators. Modeling software components and their behavior reduces the need for GPL components and liberates developers from implementation details. However, some components still require manual implementations or the integration of legacy code. As components models are tied to implementations by MontiArcAutomaton convention, architectures containing components with platform-specific implementations (PSIs) are tied to specific platforms as well. This poses challenges when generating PSIs from models.

We present a mechanism implemented in MontiArcAutomaton to enable modeling of logical, platform-independent C&C architectures and their transformation into PSIs for different platforms. This mechanism relies on a combination of model and code libraries as well as an application specific configuration that regulates the transition from models to PSIs. To illustrate the toolchain and its benefits, the next section introduces the MontiArcAutomaton modeling language and framework (Sect. 2). Afterwards, Sect. 3 explains the transformation toolchain itself and illustrates its application. Section 4 discusses related work and Sect. 5 concludes this contribution with an outlook on future work.

## 2  MontiArcAutomaton

MontiArcAutomaton [19,20] is a modeling language family and framework for generative MDE of robotics applications. Logical architectures are modeled as the hierarchical composition of components that provide the system's functionality. Components posses a stable interface comprising their type, configuration parameters, generic type parameters, and sets of typed input ports and output ports. A component is either atomic or composed. Atomic components specify behavior directly. The behavior of a composed component emerges from the interaction of its subcomponents. Components interact by sending and receiving messages over directed connectors between their ports. The types of ports are defined via class diagrams (CDs) or a GPL. Encapsulation of components with stable interfaces facilitates logically distributed development and physically distributed computation models. It enables component composition independent of their behavior description. MontiArcAutomaton exploits this encapsulation mechanism to allow the embedding of behavior modeling languages into atomic components [20]. Component developers may use the most suitable behavior description languages instead of GPLs.

MontiArcAutomaton is developed with the domain-specific language workbench MontiCore [13] which provides frameworks for language integration [14,26] and code generator development [23]. MontiCore languages are textual and defined by context free grammars with additional well-formedness rules. From these grammars, MontiCore generates infrastructure to parse complying models into their corresponding abstract syntax tree (AST). MontiCore supports language inheritance, language embedding, and language aggregation (referencing and using models from other languages) [14,26] to compose new languages from existing ones and MontiArcAutomaton uses all three mechanisms: it extends the MontiArc [11] ADL, component behavior languages are embedded into the base ADL, and port types may use UML/P [22] CD models. The MontiCore code generation framework facilitates development of code generators using the FreeMarker[3] template engine to generate code from ASTs and code templates written in a target language [18,23]. MontiArcAutomaton comprises modeling languages, code generators, generator composition mechanisms, model-transformations, language integration support, and libraries.

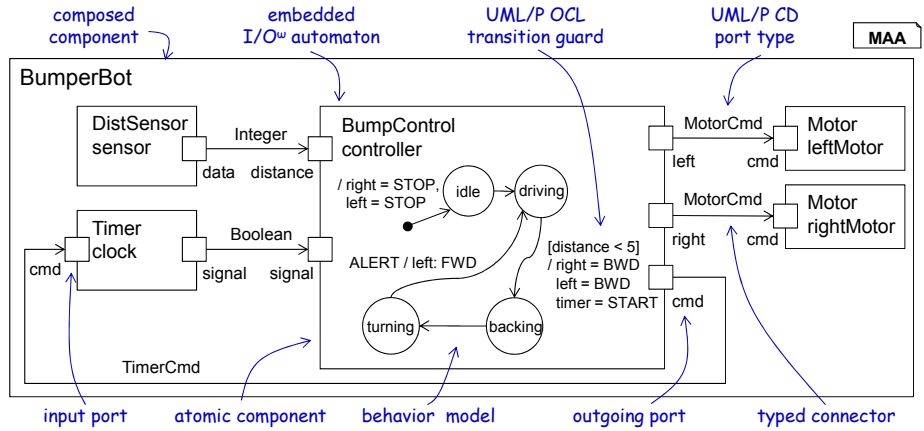

**Fig. 1.** Platform-independent software architecture of the composed component type `BumperBot` with five subcomponent instances.

Consider a robot that comprises a distance sensor to measure the distance to the closest obstacle ahead and motors to control its left and right wheel. The robot drives forward until it approaches an obstacle, then backs up, rotates, and continues to drive forward. Figure 1 depicts the logical software architecture of this robot which consists of the composed component `BumperBot` with five subcomponent instances: `sensor` of type `DistSensor`, `clock` of type `Timer`, `controller` of type `BumpControl`, and two instances `leftMotor` and `rightMotor` of type `Motor`. The subcomponent `sensor` has the single outgoing port `data` of type `Integer`, which is connected to the incoming port `distance` of `controller`. Based on the inputs received, the controller sends messages of type `MotorCmd` to the motors. This type is defined in a CD. The behavior of `controller` is modeled as an automaton following the I/O$^\omega$ automaton paradigm [17,21].

Executable code for the C&C architecture of the system requires some platform dependent component implementations. To execute the system on a Lego

---

[3] Website of the FreeMarker Java template engine: http://freemarker.org/.

NXT robot using the Lego Java Operating System (leJOS)[4] the component instances `leftMotor` and `rightMotor` require Java wrappers for the leJOS API. Executing the same system on a NXT robot using the Robot Operating System (ROS)[5] requires a Python implementation controlling ROS nodes. These platform specific components cannot easily be modeled and are among existing legacy components examples for the need of integrating GPL code in MDE.

## 3 Platform-Independent Model and Multi-Platform Code

To facilitate reuse of the same logical architecture model with different platforms, it is favorable to postpone commitment to a specific platform as long as possible. With MontiArcAutomaton this commitment is expressed as *binding* component instances to PSIs. We distinguish two kinds of components: *fully modeled components* are composed components or atomic components with an embedded behavior model. *Abstract components* are atomic components without a behavior model. The interfaces of abstract component types may refer only to types provided by the MontiArc type system and types defined in CDs. The port types depicted in Fig. 1 are such types. Fully modeled components require no binding as their implementation is generated by the combination of code generators for component structure and behavior.

**Integrating existing code:** Abstract components require GPL behavior implementations compatible with the generated code of the surrounding architecture. Integration of generated code with manual implementations can follow different patterns (e.g., generation gap [6] or delegation [8]). MontiArcAutomaton does not prescribe a pattern. Instead, MontiArcAutomaton code generators specify which *runtime environment* (RTE) they are compatible with. Such a RTE may employ appropriate patterns to integrate generated and manually implemented code, define how communication between components and scheduling are realized, and contain common domain functionality [24]. Technical details and requirements for the integrated code are RTE specific. Our RTE for Java component implementations defines an abstract class `Component` and factories [8] which enable utilization of the generation gap pattern. The code generator transforms component models into subclasses of `Component`, which realize the component behavior. For abstract components, the generator only creates the according factory and expects the component developer to provide an according component implementation in the RTE's GPL Java, i.e., to bind the component model to a PSI. A manual binding is error-prone and requires knowledge of implementation details of the generated code. Modeling the binding reduces these "accidental complexities" [7].

**Model and code libraries:** Enabling component developers to efficiently develop software architectures with abstract components requires to enable component and component implementation reuse. MontiArcAutomaton therefore distinguishes platform-independent *model libraries* and platform-specific *code libraries*. Model libraries contain fully modeled components, abstract components, and CDs. Code libraries contain component behavior implementations and port types formulated in a GPL. Furthermore, each code library contains a library properties model, describing the RTE of the contained implementations and component types each implementation conforms to. This is necessary to ensure compatibility of the generated and provided implementations for different RTEs.

---

[4] Website of the Lego Java Operating System (leJOS): http://www.lejos.org/.

[5] Website of the Robot Operating System (ROS): http://www.ros.org/.

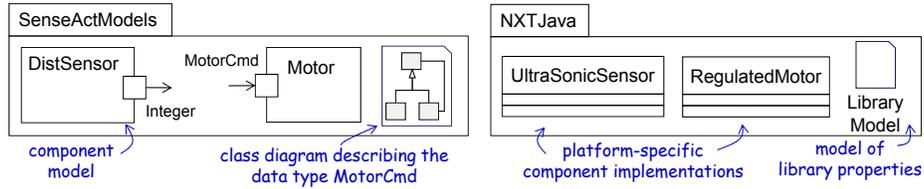

**Fig. 2.** Model library and code library used with the `BumperBot` software architecture.

The left part of Fig. 2 shows the model library `SenseActModels` used by the platform-independent `BumperBot` architecture depicted in Fig. 1. The right part shows a corresponding code library. The model library contains the abstract component models `DistSensor`, `Motor` and a CD modeling the data type `MotorCmd` used by component `Motor`. The `NXTJava` code library contains PSIs and a library properties model which describes the RTE `UltraSonicSensor` and `RegulatedMotor` are compatible with.

**Binding PSIs:** Retaining platform-independent architectures prohibits to model component binding in the logical architecture itself. Instead, MontiArcAutomaton applications may provide application configuration models. These describe the selected code generators and binding information. A binding describes a mapping of component instances of the architecture model to implementations. The mapping augments the architecture's AST before any code is generated and thus can be reused with arbitrary generator combinations.

The MontiArcAutomaton generator toolchain parses the application configuration and passes the binding information to a transformation which adds information about component implementations to the architecture. The generation framework considers this information and, e.g., generates factories instantiating the bound implementations accordingly.

Listing 1 shows the application configuration used to bind component instances `sensor`, `leftMotor`, `rightMotor`, and `clock`. First, the required implementation library is imported (ll. 1). Model libraries are imported by the architecture and made available to the application configuration. Afterwards (l. 4) code generators are selected. The generators declare which runtime environments they are compatible with and thus restrict which implementations can be bound. Finally, ll. 5-9 describe the actual bindings of the application. Here, component instances, identified by the name between `map` and `to`, are mapped to imported implementations, identified by the name after `to`. Please note that the two instances of the component motor are mapped to the same implementations `RegulatedMotor`.

```
                                                    ApplicationConfiguration
1  import NXTJava.*;
2
3  application NXTJavaBumperBot {
4    generators ComponentJava, AutomatonJava, CDJava;
5    bindings
6      map BumperBot.sensor       to UltraSonicSensor,
7      map BumperBot.leftMotor    to RegulatedMotor,
8      map BumperBot.rightMotor   to RegulatedMotor,
9      map BumperBot.clock        to JavaTimer;
10 }
```

**Listing 1.** Application configuration model for the `BumperBot` selecting code generators and binding component instances sensor, leftMotor, rightMotor, and clock.

Application configuration models are checked at design time whether all components are bound and whether the binding is compatible by reading the libraries' property models, which map the contained implementations to component types.

**Implementation in MontiArcAutomaton:** Figure 3 illustrates how the MontiArcAutomaton code generation framework integrates applications, code generators, libraries, and transformations of platform-independent architecture models into PSIs. The `GenerationTool` parses architecture and application models, which reference model libraries and code libraries, respectively. The result is passed to the `BindingTransformation` which augments the architecture before code generation. Architecture AST, binding, and imported libraries are passed to the `BindingTransformation` which transforms the AST accordingly. With the transformed AST, the `GenerationTool` starts the `GeneratorOrchestration` process which instantiates and executes the selected code generators as selected. Both, code library and code generators have to comply to the same runtime environment (RTE) to ensure an executable implementation of the architecture. The RTE provides interfaces manually implemented and generated components have to implement to ensure compatibility. Data types are translated into PSIs using the selected CD generator, which maps the basic types of the MontiArc type system onto platform-specific representations.

With help of the MontiArcAutomaton transformation toolchain, application configuration, and libraries the logical `BumperBot` architecture (Fig. 1) can be transformed into an intermediate platform-specific architecture where the subcomponents sensor, leftMotor, rightMotor, and clock are bound to PSIs. This resulting software architecture is passed to the code generation framework and ultimately transformed into implementations executable on robotic platforms.

An excerpt of the resulting implementations for two different platforms is shown in Fig. 4. The left panel shows the project structure of the `BumperBot` application containing two application configuration models. The first maps sensor, leftMotor, rightMotor and clock to Java implementations based on leJOS, the second maps them to Python implementations based on ROS. The bottom panels show part of the generated implementations for component `BumperBot` where subcomponents are instantiated. The leJOS Java implementation uses the implementations `UltraSonicSensor` and `RegulatedMotor` and the ROS python implementation uses `RangeSensor` and `JointMotor` as defined in the respective application configuration models (depicted in the top panels).

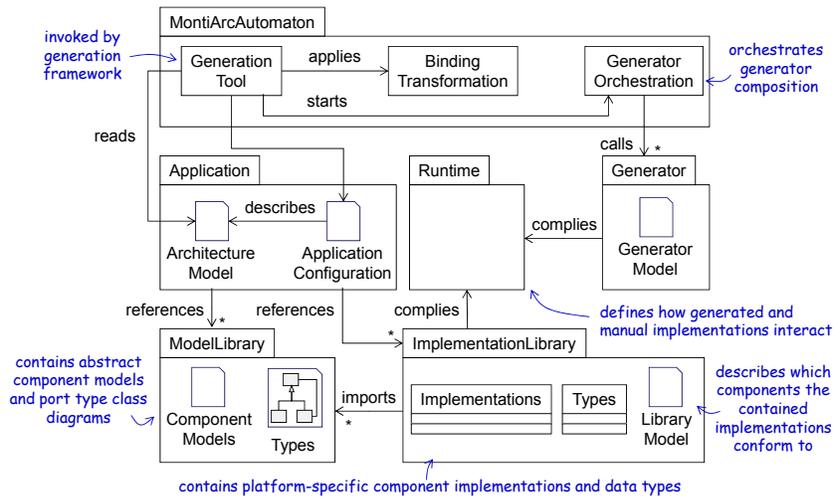

**Fig. 3.** Elements of the MontiArcAutomaton transformation toolchain and their dependencies.

## 4 Related Work

Related approaches are toolchains enabling platform-independent modeling and automated creation of source code implementations — especially ADL frameworks with code creation capabilities, e.g., the Architecture Analysis & Design Language [5] (AADL), AutoFocus [12], Simulink [25], and SysML [27].

AADL is modeling language for systems consisting of software components and hardware components. While AADL models could be subjected to late binding as well, AADL architectures models component implementations explicitly – thus hampering reuse. We are not aware of an integrated binding modeling language and framework for AADL. AutoFocus is a C&C ADL and modeling tool for the development of distributed systems based on the semantics of Focus [3]. Behavior is modeled as state transition diagrams similar to $I/O^\omega$ automata. In contrast to MontiArcAutomaton, AutoFocus lacks a distinction between component types and instances. This prohibits component reuse by instantiation and bindings as introduced above. MathWorks Simulink features a block diagram language to describe of components and connectors. Stateflow[6] extends blocks with state transition diagrams. Simulink relies on M2T code generation without intermediate model transformations. SysML is a set of modeling languages based on a subset of extended UML [10]. The SysML language for internal block diagrams resembles MontiArcAutomaton and component behavior can be modeled with state machine diagrams, thus SysML enables to express architectures similar to MontiArcAutomaton. Modeling with SysML focuses on the requirements phase and thus provides "only models on the PIM level" [9]. In most approaches manually written code (if required) is typically integrated after code generation.

---

[6] Website of Stateflow: http://www.mathworks.de/products/stateflow/.

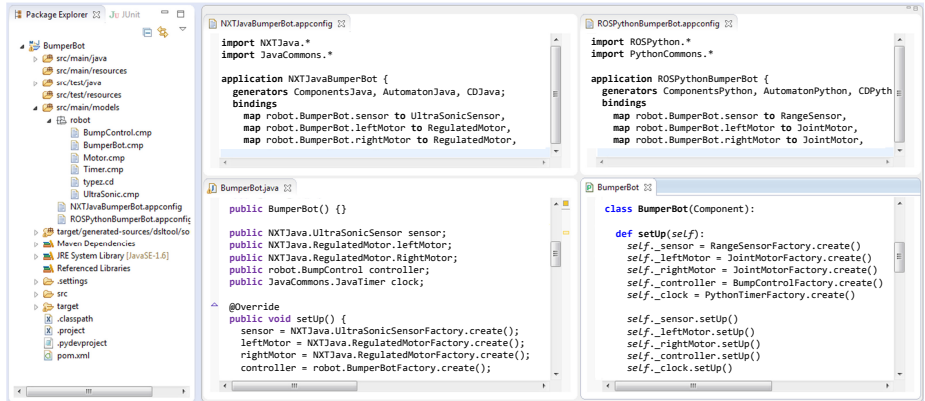

**Fig. 4.** Application configuration models and generated implementations for execution of the logical `BumperBot` architecture on two different platforms.

While we propose a binary notion of platform-independence compared to a continuous notion where "abstract platforms" [1] may add and refine platform-properties, e.g., an abstract-platform for the `BumperBot` could describe that it requires two motors. It is an interesting future work to evaluate these differences.

Other approaches to transform PIMs into PSIs focus different issues: the authors of [4], for example, transform platform-independent statecharts with real-time properties into PSMs via complex model analysis. Such languages and transformations are beyond the scope of this contribution.

## 5 Discussion and Conclusion

We presented a model-driven integrated, automated transformation toolchain, modeling languages, and library concepts for the transformation of platform-independent C&C software architectures into PSIs for multiple platforms. This transformation is defined as the binding of subcomponents to platform-specific component implementations. Abstract components are provided in model libraries while their implementations are provided in platform specific code libraries. To separate binding information for the architecture, we extended MontiArcAutomaton's application configuration modeling language to contain bindings. This separation enables reuse of logical architecture models with different source code implementations without modifications to the software architecture

Currently, bindings specify unconditional mappings. Different distribution scenarios might require to bind components under certain conditions (e.g., target platform properties). An extension of the application configuration language with conditions is easily possible due to MontiCore's language integration mechanisms. We currently explore different notions of interface compatibility as it might be feasible to bind components where a port's type might be a subtype of the abstract component's respective port. Another notion of interface compatibility is, that the replacing component extends the component of the replaced component instance in the sense of component inheritance [11]. While interface compatibility ensures syntactic well-formedness, it does not ensure that bound component implementations behave similarly. Securing this could be achieved by employing component behavior contracts. We are working on such mechanisms based on assumptions and guarantees [2].

Overall the MontiArcAutomaton toolchain integrates transformations and code generation seamlessly and enables easy reuse of the same software architecture on different platforms. In the future we plan to work on the issues mentioned above and evaluation of the toolchain.